\documentclass[journal=jacsat, manuscript=article]{achemso}
\usepackage{ulem}

\usepackage{etoolbox}

\makeatletter
\renewcommand*{\acs@author@fnsymbol@symbol}[1]{
    \ifcase #1 *\or
    1\or
    2\or
    3\or
    4\or
    5\or
    6\or
    7\or
    8\or
    9\or
    10
    \fi
}
        
\renewcommand*\acs@contact@details{
    {\sffamily *\,E-mail: \acs@email@list }%
    \acs@number@list
}           

\patchcmd{\acs@address@list@auxii}
{\acs@author@fnsymbol{\acs@affil@marker@cnt}}
{\textsuperscript{\acs@author@fnsymbol{\acs@affil@marker@cnt}}}
{}{}

\patchcmd{\acs@address@list@auxii}
{{\acs@author@fnsymbol{\acs@affil@marker@cnt}\@nameuse{@altaffil@\@roman\@tempcnta}\par}}
{{\textsuperscript{\acs@author@fnsymbol{\acs@affil@marker@cnt}}\@nameuse{@altaffil@\@roman\@tempcnta}\par}}
{}{}        

\makeatother    

\author{Tianbo Li}
\affiliation[sail]{SEA AI Lab, Singapore}
\author{Min Lin}
\affiliation[sail]{SEA AI Lab, Singapore}
\author{Stephen Dale}
\affiliation[nus]{Institute for Functional Intelligent Materials, National University of Singapore}
\author{Zekun Shi}
\affiliation[sail]{SEA AI Lab, Singapore}
\alsoaffiliation[nus2]{School of Computing, National University of Singapore}
\author{A. H. Castro Neto}
\affiliation[nus]{Institute for Functional Intelligent Materials, National University of Singapore}
\alsoaffiliation[2d]{Centre for Advanced 2D Materials, National University of Singapore}
\author{Kostya S. Novoselov}
\affiliation[nus]{Institute for Functional Intelligent Materials, National University of Singapore}
\alsoaffiliation[2d]{Centre for Advanced 2D Materials, National University of Singapore}
\author{Giovanni Vignale}
\affiliation[nus]{Institute for Functional Intelligent Materials, National University of Singapore}
\email{vgnl.g@nus.edu.sg}

\SectionNumbersOn
\title{Diagonalization without Diagonalization: \\A Direct Optimization Approach for\\ Solid-State Density Functional Theory}

\abbreviations{DFT}

\usepackage{framed}
\usepackage{graphicx}
\usepackage{dcolumn}
\usepackage{amsmath}
\usepackage{amssymb}
\usepackage{color}
\usepackage{amsmath,verbatim,moreverb,bm}
\usepackage{subcaption}
\usepackage{natbib}
\usepackage{chemformula}

\definecolor{mygray}{gray}{0.7}

\def\bef{\begin{framed}}
	\def\eef{\end{framed}}
\def\be{\begin{equation}}
	\def\ee{\end{equation}}
\def\ber{\begin{eqnarray}}
	\def\eer{\end{eqnarray}}

\def\rv{{\bf r}}

\def\Gv{{\bf G}}
\def\Fv{{\bf F}}

\def\Cv{{\bf C}}

\def\Iv{{\bf I}}

\def\Vv{{\bf V}}

\def\Xv{{\bf X}}
\def\Yv{{\bf Y}}
\def\kv{{\bf k}}

\def\nn{\nonumber}

\usepackage[colorlinks = true,
linkcolor = blue,
urlcolor  = blue,
citecolor = blue,
anchorcolor = blue]{hyperref}

\begin{document}
	
    \abstract{
    We present a novel approach to address the challenges of variable occupation numbers in direct optimization of density functional theory (DFT). By parameterizing both the eigenfunctions and the occupation matrix, our method minimizes the free energy with respect to these parameters. As the stationary conditions require the occupation matrix and the Kohn-Sham Hamiltonian to be simultaneously diagonalizable, this leads to the concept of ``self-diagonalization,'' where, by assuming a diagonal occupation matrix without loss of generality, the Hamiltonian matrix naturally becomes diagonal at stationary points. Our method incorporates physical constraints on both the eigenfunctions and the occupations into the parameterization, transforming the constrained optimization into an fully differentiable unconstrained problem, which is solvable via gradient descent. Implemented in JAX, our method was tested on aluminum and silicon, confirming that it achieves efficient self-diagonalization, produces the correct Fermi-Dirac distribution of the occupation numbers and yields band structures consistent with those obtained with SCF methods in Quantum Espresso.

    }

    \newpage
    \section{Introduction}
    The electronic structure is fundamental for predicting a material's properties, as it determines how electrons are distributed across energy bands, especially near the Fermi level, as well as the topology of these bands, which influences conductivity, reactivity, optical properties, and other key characteristics. Typically, Kohn-Sham density functional theory (KS-DFT) combined with the self-consistent field (SCF) method is used to solve for the electronic structure. However, SCF computation can fail in certain challenging cases, such as when there are multiple degenerate or near-degenerate states at the Fermi level in metals or when the gap between the highest occupied molecular orbital (HOMO) and the lowest unoccupied molecular orbital (LUMO) is very small in molecules. \cite{mrovec2021diagonalization, ivanov2021method, levi2020variational, levi2020variational2, yao2024enhancing} In these scenarios, SCF can result in band/orbital reordering between iterations, leading to oscillations and poor convergence. \cite{schlegel1991you, daniels2000converging, jorgensen2021effectiveness, ren2022impacts} To address these issues, direct optimization methods have been developed, offering more reliable convergence by bypassing the iterative nature of SCF. \cite{marzari1997ensemble, marzari1999thermal, ivanov2021direct, ivanov2021method, levi2020variational, freysoldt2009direct, mrovec2021diagonalization, pham2024direct, wu2005direct, yang2006constrained, yang2007trust} 
    For example, a series of works by \cite{yang2006constrained, yang2007trust} investigate  direct optimization methods for solving Kohn-Sham density function functional theory and demonstrate that this approach can be more stable than the traditional SCF technique.
    Additionally, \citet{cances2021convergence} systematically analyze both SCF and direct optimization methods, concluding that in condensed-matter physics, especially when using plane wave bases, direct optimization may offer greater efficiency and robustness.
    Moreover, the rise of machine learning-enhanced DFT has brought renewed attention to direct optimization, as it aligns well with the training processes of machine learning models. Despite these advantages, direct optimization still encounters challenges in efficiently managing occupation numbers. \cite{cances2021convergence} The complexity stems from constraints such as the Pauli exclusion principle, electron conservation, and Fermion statistics, which become particularly intricate when combined with energy minimization. Therefore, accurately and efficiently handling occupation numbers remains a critical challenge in direct optimization methods.
    Early works did not vary the occupations, instead fixing them at a ``reasonable'' approximation of metallic behavior—fully occupied well below the Fermi level, half-occupied near the Fermi level, and unoccupied elsewhere.\cite{grumbach1994ab} A seminal piece of literature on solving the occupation issue in direct optimization of DFT is a dual-loop direct optimization technique proposed by \citet{marzari1997ensemble}, which optimizes the free energy of a system with respect to the orbitals in the outer loop, and optimizes of the occupations of those orbitals in the inner loop. 
    This method demonstrated robust convergence, and the direct optimization of the occupations was subsequently incorporated into SCF calculations to enhance their stability. As a result, it has become one of the most widely used examples of direct optimization in DFT \cite{marzari1999thermal}. Some works recognize the double loop optimization of orbitals and occupations to be problematic and instead modify both the orbitals and the occupations within the same loop,\cite{ismail2000new} sometimes with a preconditioner designed to stabilise convergence.\cite{ivanov2021method, levi2020variational, levi2020variational2} A different way of creating a single loop approach is to generate a pseudo-Hamiltonian matrix, which is a function of both the orbitals and the occupations, as in \citet{freysoldt2009direct}, the free-energy is then optimized with respect to the pseudo-Hamiltonian while enforcing the diagonal condition at each step.

    In this paper, we address the challenge of managing occupation numbers within the framework of direct optimization in density functional theory (DFT). As in previous methodologies, we parameterize the eigenfunctions and occupation matrix and aim to minimize the free energy with respect to these variational parameters. Our key observation is that, at the stationary point of this optimization problem, the occupation matrix and the Kohn-Sham Hamiltonian must be simultaneously diagonalizable, in accordance with Liouville's theorem. Due to the unitary invariance of the free energy, there are infinitely many stationary points where the free energy is stationary with respect to infinitesimal variations of the wave function. However, all these stationary points are unitarily equivalent to the Kohn-Sham formulation, meaning there always exists a unitary transformation that can transform the free energy minimization into a Kohn-Sham equation, simultaneously diagonalizing both the occupation matrix and the Kohn-Sham Hamiltonian. This insight led us to assume that if the occupation matrix is constrained to be diagonal, then, at the stationary point of free energy, the Kohn-Sham Hamiltonian will also become diagonal. We refer to this phenomenon as the ``self-diagonalization'' of the Kohn-Sham Hamiltonian. This assumption speeds up the computation of the energy by an order of magnitude (see Fig.~(\ref{fig:scaling})) since the free energy only depends on the diagonal elements of the Hamiltonian.

    Our approach to parameterization  differs significantly from previous methods. We hard-wire all constraints of the wave functions and occupation matrix into the parameterization, transforming the constrained optimization problem into an unconstrained one. For the orthonormality constraint of the wave functions, we utilize a QR decomposition approach. This method is applicable to rectangular matrices, unlike conventional matrix exponential and Cayley transformations, which are limited to square matrices. This makes our method more suitable for solid-state DFT using plane waves, where the number of basis functions greatly exceeds the number of orbitals so that the latter are naturally represented by elongated rectangular matrices. For the occupation matrix parameterization, we propose a novel approach that ensures the occupation numbers always satisfy the Pauli exclusion principle, charge conservation, and Hermiticity. 
    Additionally, it offers better differentiability, allowing the gradient of the variational parameters to be easily obtained using automatic differentiation frameworks. We implemented our algorithm using JAX and tested it on real systems, such as aluminum and silicon. Our numerical experiments confirmed that optimizing the free energy with the diagonal assumption for the occupation matrix leads to the self-diagonalization of the Hamiltonian matrix, and the corresponding occupation matrix exhibits a Fermi-Dirac distribution. This approach also produces the same band structure as the SCF method implemented in Quantum Espresso, validating the correctness of our proposed method.

    This paper is organized as follows: Section \ref{sec:theory} provides the theoretical formulation of free energy minimization in DFT. We also establish the stationary condition, illustrate its connections to the self-consistent Kohn-Sham equations, and provide the theoretical basis for the ``self-diagonalization'' of the Hamiltonian matrix under the assumption of a diagonal occupation matrix. Section \ref{sec:implementation} describes our parameterization of the eigenfunctions and the occupation matrix, as well as the optimization algorithm. Section \ref{sec:results} presents four numerical tests that validate the effectiveness of the proposed method.


\newpage
\section{Theory}
\label{sec:theory}

In this section, we review the fundamental concepts and theoretical results related to free energy minimization in DFT and its connection to the self-consistent Kohn-Sham equations. We derive the stationary conditions for free energy minimization with respect to the eigenfunctions and the occupation matrix, leading to the condition established by the Liouville theorem. Finally, we propose an efficient strategy for optimizing the free energy: by restricting the occupation matrix to a diagonal form and minimizing the free energy, the Kohn-Sham Hamiltonian naturally becomes diagonal, a process we refer to as self-diagonalization.



\subsection{Density Functional Theory}
The starting point of our approach is the \citet{hohenberg1964inhomogeneous}'s density functional theory (DFT) as generalized by \citet{mermin1965thermal} to thermal ensembles. According to this the  equilibrium density of the many-electron system is obtained by minimizing the free energy $A[n]$, regarded as a functional of the electronic density $n(\rv)$:
    \be\label{Omega}
    A[n]=T_s[n]+E_{Hxc}[n]+\int V_{ext}(\rv)n(\rv)d\rv -T S[n]\,,
    \ee
where $T_s[n]$ is the non-interacting kinetic energy functional, $E_{Hxc}[n]$ is the sum of Hartree and exchange-correlation (xc) energy functionals, $S[n]$ is the entropy functional and $T$ is the temperature.

A key assumption of density functional theory is that the minimizer of $A[n]$, i.e., the exact density of the electronic system, can be obtained as the equilibrium density of an auxiliary {\it non-interacting} system, known as the {\it Kohn-Sham system}, which is described by the non-interacting single particle Hamiltonian
\be\label{HS}
    \hat H_s = - \frac{\nabla_\rv ^2}{2}+ V_{Hxc}(\rv) +V_{ext}(\rv)
\ee
where $V_{Hxc}(\rv) \equiv \frac{\delta E_{Hxc}[n]}{\delta n(\rv)}$ is the local one-body potential constructed as the functional derivative of the $E_{Hxc}$ functional with respect to the density.  This potential is itself a functional of the density and can be computed from a suitable approximation to the xc energy functional.  


\subsection{Occupation Matrix}
\label{sec:occupation}
The occupation matrix is crucial for systems with partially occupied electronic states, particularly in metals  at finite temperatures, where electronic states are neither fully occupied nor completely unoccupied. The eigenvalues of the occupation matrix determine the occupations of the eigenstates of the Kohn-Sham Hamiltonian $\hat H_s$. For periodic systems, which are the focus here, let $\psi_i(\kv,\rv)$ represent a sufficiently large set of orthonormal eigenfunctions, collectively denoted by $\{\psi\}$. These eigenfunctions are characterized by a band index $i=1,..., I$, where $I$ is larger than the number of electrons $N$ per unit cell, and a Bloch wave vector $\kv$. Each band contains $K$ wave vectors, where $K$ corresponds to the number of unit cells over which Born von-Karman periodic boundary conditions are applied.  We will refer to this set of orbitals as \textit{potentially occupied orbitals}. 

Following the works of \citet{marzari1997ensemble}, \citet{ivanov2021method} and \citet{gonze2024variational}, the actual occupations are specified occupation matrix $\hat \gamma$ with matrix elements $\gamma_{ij}(\kv)$ such that the density takes the form
	\be\label{Density}
	n(\rv)=\sum_{ij,\kv}\gamma_{ij}(\kv)\psi^*_j(\kv,\rv)\psi_i(\kv,\rv)\,.
	\ee
In this equation, the  sum runs over $\kv$ in the first Brillouin zone, and
	\be\label{Orthonormality}
	\int \psi^*_j(\kv,\rv)  \psi_i(\kv',\rv) d\rv = \delta_{\kv,\kv'}\delta_{ij}\,.
	\ee
The occupation matrix $\hat \gamma$ must have the following properties: 
\begin{enumerate}
    \item \textit{Hermiticity}. The occupation matrix $\hat \gamma$ must be Hermitian, meaning it satisfies
    \be\label{occ:hermitian}
        \hat \gamma = \hat \gamma ^\dagger.
    \ee
    This condition ensures that the eigenvalues of $\hat \gamma$ are real.
    \item \textit{Pauli's exclusion principle}. The eigenvalues of $\hat \gamma$, denoted as $\{f_i\}$, are all comprised between $0$ and $1$:
	\be\label{occ:range}
	0 \leqslant  f_i \leqslant 1,  \quad \forall i = 1, \dots, N;
	\ee 
\item \textit{Charge Conservation}. The trace of $\hat\gamma$ (summed over $\kv$) equals the total number of electrons,
	\be\label{occ:trace}
	   \sum_{\kv}{\rm Tr}\{\hat \gamma(\kv)\}=\sum_{i,\kv}\gamma_{ii}(\kv)=NK;
	\ee
\end{enumerate}
 


\subsection{Unitary Invariance}

With the occupation matrix $\hat\gamma$, the free energy can be represented as a functional of $\hat\gamma$ and $\{\psi\}$ as follows:
	\be\label{Omega2}
	A[\hat\gamma,\{\psi\}]=E[\hat\gamma,\{\psi\}] - TS[\hat\gamma]
	\ee
	where
	\ber\label{Energy}
	E[\hat\gamma,\{\psi\}]=-\frac{1}{2}\sum_{ij,\kv}\gamma_{ij}(\kv)\int \psi^*_j(\kv,\rv)\nabla_{\rv}^2\psi_i(\kv,\rv)+E_{Hxc}[n]+\int V_{ext}(\rv) n(\rv) d\rv,
	\eer
	is the energy and the density is given by Eq.~(\ref{Density}).   $S[\hat\gamma]$ is the entropy of a noninteracting system of fermions with occupation matrix $\hat \gamma$, i.e.,
	\be\label{Entropy}
	S[\hat\gamma] = -\sum_{\kv}{\rm Tr}\left\{\hat\gamma(\kv) \ln \hat\gamma(\kv) +[\hat1-\hat\gamma(\kv)]\ln[\hat1-\hat\gamma(\kv)]\right\}\,.
	\ee
    
    This function is invariant with respect to a unitary transformation of the potentially occupied orbital among themselves, combined with the corresponding transformation of the occupation matrix, i.e., the transformation
	\be
	\psi_i(\kv,\rv) \to \sum_j U_{ij}(\kv) \psi_j(\kv,\rv)\,,~~~~\gamma_{ij}(\kv) \to  \sum_{i'j'}U_{ii'}(\kv)\gamma_{i'j'}(\kv) U^\dagger_{j'j}(\kv)\,,
	\ee
    leaves the density, the energy, and in general all the physical properties of the system unchanged.
    This implies that the solution of the minimization problem cannot be unique, since we can always apply a unitary transformation that does not change the physical properties. 

    \subsection{Stationary Conditions}
    To minimize  the free energy $A$ under the orthonormality constraint on $\{ \psi \}$ and the trace constraint of $\hat\gamma$ (Eq.~\eqref{occ:trace}), one typically employs the Lagrange multiplier method. The Lagrangian can be written as
    \be
        \mathcal{L} = A[\hat\gamma,\{\psi\}] - \sum_{ij, \kv} \lambda_{ij}(\kv) \left( \int \psi^*_j(\kv,\rv)  \psi_i(\kv,\rv) d\rv - \delta_{ij} \right) - \mu \left(\sum_{\kv}{\rm Tr}\{\hat \gamma(\kv)\} - NK\right)
    \ee
    where $\lambda_{ij}(\kv)$ is a Hermitian matrix of Lagrange multipliers (collectively denoted by $\hat\lambda(\kv)$) that enforces the orthonormality constraint at each $\kv$, and $\mu$ is the Lagrange multiplier for the trace constraint, 
    commonly referred to as the chemical potential.  At the stationary point of $A$ (defined by Eq. \eqref{Omega2} with the above-specified constraints), the system must satisfy the following conditions with respect to both the occupation matrix $\hat \gamma$ and eigenfunctions $\{\psi\}$:
    \be
        \dfrac{\delta  \mathcal{L}}{\delta \psi_i^*(\rv)} = 0, \quad\quad  \dfrac{\delta  \mathcal{L}}{\delta \gamma_{ji}(\rv)} = 0.
    \ee
    In the following, we examine how these stationary conditions connect to the conventional Kohn-Sham DFT framework. We also focus on the critical relationship between the occupation matrix and the Kohn-Sham Hamiltonian, a connection demanded by the Liouville theorem.
    
    \subsubsection{Stationary Condition for $\psi$ }
    Requiring that $A$ is stationary with respect to infinitesimal variation of $\psi_i^*(\rv)$, i.e. $\frac{\delta \mathcal{L}}{\delta \psi_i^*(\rv)} = 0$,  yields the equation
    \be\label{KSE}
    \sum_j \gamma_{ij}(\kv) \hat H_s\psi_j(\kv,\rv) =  \sum_j \lambda_{ij}(\kv)  \psi_j(\kv,\rv)\,,
    \ee
    If $\hat \gamma$ is invertible, which holds at finite temperature, we can define:
    \be\label{eq:epsgammalambda}
    \hat h(\kv) = ({\hat\gamma(\kv)})^{-1} \hat\lambda(\kv)\,,
    \ee
    with $h_{ij}(\kv)$ represents the matrix element at a given $\kv$. With this substitution, the equation simplifies to:
    \be
          \hat H_s\psi_i(\kv,\rv) =  \sum_j h_{ij}(\kv)  \psi_j(\kv,\rv)\,,
    \ee 
    At the equilibrium the matrix elements $h_{ij}(\kv)$ are expressed as:
    \be\label{eq:hamil_matrix}
        h_{ij}(\kv) = \int \psi^*_j(\kv,\rv) H_s\psi_i(\kv,\rv) d\rv
    \ee
    which is commonly referred to as the \textit{Kohn-Sham Hamiltonian matrix}.
    
    Furthermore, $\hat{h}(\kv)$ can be brought to diagonal form by a unitary transformation, such that $h_{ij}(\kv)\to \varepsilon_i(\kv) \delta_{ij}$. In this case, this stationary condition reduces to the Kohn-Sham Equation:
    \be
         \hat H_s\psi_i(\kv,\rv) =  \varepsilon_{i}(\kv)  \psi_i(\kv,\rv)\,,
    \ee
    This establishes the fact that the set of optimal orbitals $\{\psi\}$ is \textit{unitarily equivalent} to the set of eigenfunctions of the Kohn-Sham Hamiltonian, with eigenvalues $\varepsilon_i(\kv)$. In other words, there always exists a unitary transformation that can convert the optimal $\{ \psi \}$ that extremize $A$ into the solutions of the Kohn-Sham equation.
    
    \subsubsection{Stationary Condition for $\hat\gamma$}
    \label{sec:liouville}
 Requiring that $A$ is stationary with respect to infinitesimal variation of $\gamma_{ij}(\kv)$, i.e., $\frac{\delta  \mathcal{L}}{\delta \gamma_{ji}(\rv)} = 0$  yields the equation (by differentiation of Equation~\ref{Omega2})
	\be
	T\frac{\delta S[\hat \gamma]}{\delta\gamma_{ji}(\kv) }=\frac{\delta E[\hat \gamma]}{\delta\gamma_{ji}(\kv)}-\mu\delta_{ij}\,,
	\ee
    where $\mu$ (chemical potential) is the Lagrange multiplier that enforces the trace constraint~(\ref{occ:trace}).
	It is not difficult to show, making use of Eq.~(\ref{Energy}) and the expression of $V_{Hxc}(\rv)$ as a functional derivative of $E_{xc}[n]$, that
	\be\label{Janak}
	\frac{\delta E[\hat \gamma]}{\delta\gamma_{ji}(\kv)} = h_{ij}(\kv)\,,
	\ee
	a result better known as Janak's theorem \cite{janak1978proof}.  Thus we have
	\be\label{StationaryF}
	T\frac{\delta S[\hat \gamma]}{\delta\gamma_{ji}(\kv) }=h_{ij}(\kv) -\mu\delta_{ij}\,.
	\ee

    \paragraph{Liouville Theorem} The unitary transformation that diagonalizes $h_{ij}(\kv)$ on the right hand side of Eq. \eqref{StationaryF} must also diagonalize the left hand side, when the stationary condition is fulfilled.\cite{landaulifshitz}
    This is possible only if both $\hat\gamma$ and $\hat h$ are diagonalized by the same transformation. 
     In other words, at the stationary point, the occupation matrix $\hat\gamma$ commutes with the Hamiltonian matrix $\hat h$, a result commonly known as the Liouville's theorem \cite{breuer2002theory}: 
    \be\label{liouville}
        \Big[\hat\gamma,  \ \hat{h} \Big] = 0.
    \ee
    
     \paragraph{Fermi-Dirac Distribution} The physical entropy functional defined in Eq.~(\ref{Entropy}) and Eq.~(\ref{StationaryF})  leads to a natural connection between the eigenvalues of $\hat \gamma$ and those of $\hat H_s$:
	\be\label{EigenvalueConnection}
	\ln \frac{1- f_{i}(\kv)}{f_i(\kv)} = \frac{\varepsilon_i - \mu}{T}\,,
	\ee
	showing that 
	\be
	f_i(\kv)= \frac{1}{e^{(\varepsilon_i(\kv)-\mu)/T}+1}\,,
	\ee 
	which is the Fermi-Dirac distribution at energy $\varepsilon_i(\kv)$, temperature $T$ and chemical potential $\mu$.  Notice that $f_i(\kv)$ is always comprised between $0$ and $1$, as it should. Different choices of the entropy functional are possible and even recommended in metallic systems when the purpose is not to describe the effect of the temperature on the system properties but to accelerate the convergence of the calculation at essentially zero temperature \cite{methfessel1989high, dos2023fermi}. However, the property $0 \leq f_i(\kv)\leq 1$ must always remain in force. 

    \subsection{Self-diagonalization of the Kohn-Sham Hamiltonian} \label{sect:self-diagonalization-ham}
    We observe that due to the unitary invariance, the stationary points of the free energy, even if it is not the true minimum, are not unique. But the stationary condition shown in Eqs. \eqref{KSE} and \eqref{liouville} guarantees that every stationary solution corresponds to a self-consistent solution of the Kohn-Sham equation, where both the Kohn-Sham Hamiltonian matrix $\hat{h}$ and the occupation matrix $\hat\gamma$ are simultaneously diagonal. Consequently, the free energy defined in Eq. \eqref{Omega2} can now be expressed solely in terms of the self-consistent eigenvalues $\varepsilon_i$ and $f_i$:
    \ber\label{ObjectiveA}
	A &=& \sum_{i,\kv}f_i(\kv)\varepsilon_i(\kv)-\int V_{Hxc}(\rv)n(\rv) d\rv+E_{Hxc}[n] \nn\\
	&-& T\sum_{i,\kv}\left\{f_i(\kv)\ln f_i(\kv)+[1-f_i(\kv)]\ln [1-f_i(\kv)]\right\}\,,
    \eer
    where the second term on the right hand side removes the interaction contribution to the Kohn-Sham eigenvalues and the third term restores the correct interaction energy.
    The self-consistent density can be written, as  
    \be
	n(\rv)=\sum_{i,\kv}f_i(\kv) |\psi_i(\kv,\rv)|^2\,,
    \ee
    where $\psi_i$ are eigenfunctions of the Kohn-Sham Hamiltonian with eigenvalues 
    \begin{equation}
    \varepsilon_i(\kv)=\int \psi^*_i(\kv,\rv) \hat H_s \psi_i(\kv,\rv) d\rv.  \label{eq:eigenvalues}
    \end{equation}


    Eq. \eqref{ObjectiveA} shows that to find the minimum of the free energy and solve the corresponding Kohn-Sham equation, we only need  the eigenvalues of the occupation matrix $\hat \gamma$ and the Hamiltonian $\hat{H}_s$. This leads to a more efficient strategy for free energy optimization: \textit{we parameterize $\hat \gamma$ such that it is always diagonal}, satisfying all the constraints in Eqs. \eqref{occ:hermitian}, \eqref{occ:range}, and \eqref{occ:trace}. As a result, according to the Liouville equation (Eq. \eqref{liouville}), the stationary conditions of the free energy forces the Kohn-Sham Hamiltonian to become diagonal, as it must commute with the diagonal occupation matrix that we parameterize. 
This approach has two significant advantages:
\begin{enumerate}
    \item Self-diagonalization of the Kohn-Sham Hamiltonian. The stationary condition for the free energy $A$ requires that the occupation matrix $\hat\gamma$ commutes with $\hat{h}$. Given that $\hat\gamma$ is already in diagonal form, the eigenstate that minimize the free energy $A$ will simultaneously diagonalize $\hat{h}$. This process, which we refer to as \textit{self-diagonalization} of the Hamiltonian, occurs naturally during the search for eigenfunctions, without the need for explicit eigendecomposition.
    \item Computational Efficiency. Since we assume the occupation matrix is diagonal and the free energy in Eq.~\eqref{ObjectiveA} depends only on the diagonal elements of $\hat h$, we can dispense with the full matrix  $\hat h$ and only store the diagonal elements of $\hat h$. This significantly reduces both the computational time and storage requirements associated with the off-diagonal elements of $\hat h$.
\end{enumerate}

    
	
    \newpage
    \section{Implementation}\label{sec:implementation}
    From the above derivation, we conclude that it is sufficient to parameterize the occupation matrix in diagonal form, i.e., in terms of its eigenvalues, while satisfying the constraints introduced in Section \ref{sec:occupation}. By minimizing the free energy with respect to both the eigenvalues of the occupation matrix and the potentially occupied orbitals, the Kohn-Sham Hamiltonian will naturally self-diagonalize. 
    This approach has three distinguishing characteristics compared to previous methods: 
    \begin{enumerate}
        \item All physical constraints—such as positivity, the Pauli exclusion principle, and the orthonormality of the orbitals—are  incorporated into the parameterizations of the occupations and potentially occupied orbitals. This allows our algorithm to transform the constrained optimization of the free energy into an unconstrained optimization problem.
        \item Our algorithm yields a diagonal Kohn-Sham Hamiltonian without explicitly performing any diagonalization operations, which reduces both computational time and storage requirements.
        \item We combine the optimization of occupations and orbitals into a single gradient descent search with a single objective. This sets our method apart from other methods in the literature which require a separate loops for occupations and orbitals.\cite{marzari1997ensemble, ivanov2021method, fattebert2022robust} 
    \end{enumerate}
    
    \subsection{Parameterizing the Potentially Occupied Orbitals}
    \label{sec:param_orbitals}
    As a first step in our optimization procedure the potentially occupied Bloch wave functions $\psi_i(\kv,\rv)$ are expanded in plane waves across the periodic crystal lattice:
	\be \label{eq:bloch}
	\psi_i(\kv,\rv) = \frac{1}{\sqrt{\Omega}}
	\sum_{\Gv}C_{\Gv,i}(\kv)\exp\left[\text{i}(\kv + \Gv)\cdot\rv\right]\,
	\ee
    where the sum runs over reciprocal lattice vecotrs $\Gv$ and $\Omega$ is the volume of the unit cell. 
    The coefficients $C_{\Gv,i} (\kv)$  are indexed by a reciprocal lattice vector $\Gv$ (row index), and a band index $i$, and are functions of  $\kv$ in the first Brillouin zone. These coefficients  are subject to an orthonormality constraint at each given $\kv$,
	\be
	\label{eq:orthogonal}
	\sum_{\Gv} C^*_{\Gv,i}(\kv) C_{\Gv,j} (\kv) = \delta_{ij}.
	\ee
	We arrange the coefficients in an elongated rectangular array $\Cv_\kv$ of dimensions $G\times I$, where $G$ is the number of basis states (plane waves) and $I$ is the number of potentially occupied bands. In this notation the orthonormality constraint takes the form
	\be
	\Cv^\dagger_\kv \cdot \Cv_\kv = {\bf I}_I\,,
	\ee
	where $\Cv^\dagger_\kv$ is a rectangular $G\times I$ matrix,   ${\bf I}_I$ is the $I\times I$ identity matrix and the dot denotes the usual matrix multiplication.
	
    In the standard KS-DFT approach the orthogonality of the Kohn-Sham orbitals is guaranteed by the hermiticity of the Kohn-Sham equation.  As we move away from KS-DFT a crucial question arises: how can we incorporate the orthogonality constraint for the coefficients $C_{\Gv,i}(\kv)$? A common method involves using unitary transformations, parameterized as  exponentials  of skew-Hermitian matrices \cite{freysoldt2009direct, levi2020variational, ivanov2021direct, ivanov2021method, mortensen2024gpaw}, to generate orthogonal sets of potentially occupied orbitals. However, the exponential transformation method can only be applied to square matrices. Therefore, it works in situations in which the number of basis functions equals the numbers of potentially occupied orbitals. In solid state applications using a plane wave basis, the number of basis functions greatly exceeds the number of bands at each $\kv$ vector. Therefore, this method becomes inefficient as we only require a few eigenstates for each $\kv$. Moreover,  The matrix exponential operation necessitates multiple eigendecompositions and matrix multiplications, which can significantly slow down the computational process. 

    
    To avoid these problems, we parameterize the orthogonality constraint for the coefficients $\Cv_\kv$ via QR decomposition. Starting with an arbitrary $G\times I$ matrix $\Xv_\kv$  the QR decomposition expresses $\Xv_\kv$ as the product of a $G\times I$ matrix with orthonormal columns (denoted by $\boldsymbol{Q}$) and an $I\times I$ upper triangular matrix (denoted by $\boldsymbol{R}$).  Discarding $\boldsymbol{R}$ we use $\boldsymbol{Q}$ as our desired orthogonal matrix $\Cv_{\kv}$. Thus, the parameterization of this the orthonormal coefficients can be expressed as follows:
	\be\label{QR1}
	\Cv_\kv = {\sf QR}[\Xv_\kv]
	\ee
	where $\Xv_\kv$ denotes the set of variational parameters that are subject to optimization and ${\sf QR}[\Xv_\kv]$ denotes the orthonormal part of the QR decomposition of the matrix $\Xv_\kv$. This  form ensures that the orthonormality of the wavefunction coefficients is maintained during the optimization process. Compared to the matrix exponential operation, QR decomposition is more efficient and can be applied to rectangular matrices. Moreover, the gradient of QR decomposition has been studied in detail \cite{roberts2020qr} and is implemented in modern automatic differentiation frameworks such as PyTorch and JAX. 
    
    \subsection{Parameterizing the Occupation Matrix} 
    Next we consider the constraints on the occupation matrix.  As discussed in the previous section we take advantage of the invariance of the free energy with respect to unitary transformations to restrict our search to diagonal occupation matrices
	\be
	\gamma_{ij}(\kv)=f_i(\kv)\delta_{ij}\,.
	\ee
	These occupations can be arranged in a square diagonal  matrix of dimension $IK \times IK$, denoted by $\Fv$, in the following manner:
	\be
	\Fv := 
	\begin{pmatrix}
        f_{1}(\kv_1) &  &  &  &  &  & \\
        & \ddots &  &  &  &  &  \\
		&  & f_{1}(\kv_K) &  &  &  &  & \\
		&  &  &  \ddots &  &  &      \\
		  &  &  &  &  f_{N}(\kv_1) & & \\
		  &  &  &  &  & \ddots &       \\
		  &  &  &  &  &  & f_{N}(\kv_K)\\
	\end{pmatrix}
	\ee
     The matrix is indexed jointly by both the orbital and the $\kv$ point. Note that, due to unitary invariance, this assumption involves no loss of generality. As we have required that $\hat\gamma$ is diagonal, then the Hamiltonian matrix $\hat h$ will  necessarily be diagonal at the solution of the optimization problem.  One might object that $\hat h$ needs not be diagonal in the degenerate subspaces of $\hat \gamma$.   In a strictly mathematical sense, Eq.~(\ref{EigenvalueConnection}) precludes this possibility, since it mandates that $\hat \gamma$ and $\hat h$ are not only simultaneously diagonal, but also simultaneously degenerate.  In practice the occupation numbers rapidly (exponentially) converge to $1$ or $0$ for states that are far from the Fermi level, even though their energies are widely different. But for  states in these occupation-degenerate subspaces it is practically irrelevant whether the Hamiltonian is diagonal or not, because their contribution to the energy is given simply by the trace of the Hamiltonian in the degenerate subspace, which is invariant under unitary rotations in the subspace. 
    

    The diagonal elements of $\Fv$ must be comprised between $0$ and $1$ and add up to $NK$ (notice that $N\leq I$).  To ensure satisfaction of this constraint we propose to the following parameterization:
	\be
	\Fv=  \text{diag}\left( \Vv \cdot \Vv^\dagger \right) \label{eq:occ_def}
	\ee
	where $\Vv$ is an $IK \times NK$ matrix of orthonormal columns which is generated by applying the QR decomposition to an arbitrary rectangular matrix $\Yv$  of  dimensions $IK \times NK$  and discarding the upper triangular part:
	\be\label{QR2}
	\Vv = {\sf QR}(\Yv)\,. 
	\ee
	Similar to Eq.~(\ref{QR1}), the elements of the matrix $\Yv$ are variational parameters subject to optimization.
	Since, by construction, we have 
	\be\label{INK}
	\Vv^\dagger \cdot \Vv = \Iv_{NK}\,,
	\ee
	where $\Iv_{NK}$ is the $NK \times NK$ identity matrix we immediately see that
	\be
	{\rm Tr}[\Fv]={\rm Tr}[\Vv\cdot\Vv^\dagger]= {\rm Tr}[\Vv^\dagger\cdot\Vv] =NK
	\ee
	as required.  In addition, the $IK\times IK$ square matrix $\tilde \Fv \equiv \Vv\cdot\Vv^\dagger$ is idempotent:
	\be
	\tilde \Fv^2 =\tilde\Fv
	\ee
	which follows immediately from Eq.~(\ref{INK}). 
	This implies that its diagonal elements, which by construction are the occupation numbers $f_{i}(\kv)$, are all comprised between $0$ and $1$.  Indeed,
	\be
	\tilde F_{i\kv,i\kv}= \sum_{\kv'}\sum_{j=1}^{I} \tilde F_{i\kv,j\kv'}\tilde F_{j\kv',i\kv} \geq \tilde F_{i\kv,i\kv}^2\,,
	\ee 
	from which the desired inequality
	\be
	0 \leq \tilde F_{i\kv,i\kv}\leq 1
	\ee
	follows immediately.

    \subsection{Algorithm}
    \label{sec:algorithm}
    The parameterization method above reformulates the constrained search problem of free energy into an unconstrained minimization problem:
	\be \label{eq:free_energy}
	\min_{\Yv,\ \Xv_{\kv}} \ E\left[\{f_{i\kv}\}(\Yv), \{\psi_{i}\}(\Xv_\kv)  \right] - TS\left[\{f_{i\kv}\}(\Yv)\right]\,,
	\ee
    with variational parameters $\Yv$ and $\Xv_\kv$ (where we have defined $f_{i\kv}\equiv f_i(\kv)$).
    This approach incorporates the orbital and occupation matrix's constraints of finite-temperature KS-DFT, as described in Section \ref{sec:occupation} and Section \ref{sec:param_orbitals}. All operations in these equations are differentiable, meaning the gradients are well-defined and can be computed using automatic differentiation frameworks such as JAX\cite{jax2018github} or PyTorch \cite{paszke2019pytorch}. Consequently, gradient-based optimizers can be employed. The gradient-based optimization process for the proposed method is summarized as follows
    \begin{enumerate}
    \item Initialize the variational parameters $\Yv$ and $\Xv_\kv$.
    \item Repeat the following steps until convergence criterion is met:
    \begin{enumerate}
        \item Compute the diagonal occupation matrix $\Fv$ via Eqs. \eqref{eq:occ_def} and \eqref{QR2};
        \item Construct the Bloch wave function of each band $\psi_i(\kv,\rv)$ via Eqs. \eqref{eq:bloch} and \eqref{QR1};
        \item Compute the free energy Eq. \eqref{eq:free_energy};
        \item Calculate the gradient of free energy with respect to the variational parameters $\Yv$ and $\Xv_\kv$;
        \item Update the parameters $\Yv$ and $\Xv_\kv$.
    \end{enumerate}
    \end{enumerate}
    Parameter updates can be carried out using steepest gradient descent, second-order methods like L-BFGS \cite{liu1989limited}, or adaptive gradient methods commonly used in deep learning, such as AdaGrad\cite{duchi2011adaptive} and Adam \cite{kingma2014adam}. In self-consistent field (SCF) calculations, preconditioning techniques are often employed to improve convergence and stability. In direct optimization, adaptive gradient methods like AdaGrad and Adam have been shown to offer similar preconditioning benefits \cite{duchi2011adaptive, das2024towards}.

    \paragraph{Complexity Analysis} 
    Our approach has two primary bottlenecks: the QR decomposition and the fast Fourier transformation (FFT) operations for the calculation of Coulomb integrals in reciprocal space. The algorithm employs two QR decompositions: one for plane wave coefficients and another for occupation numbers. The first QR decomposition for plane wave coefficients has a complexity of $\mathcal{O}(KGI^2)$, comparable to the Davidson algorithm for diagonalization in self-consistent field (SCF) methods.
    If a matrix exponential operation is used to parameterize orthonormality, the complexity increases to $\mathcal{O}(KG^3)$ since it applies only to square matrices. The QR decomposition for the occupation matrix incurs a complexity of $\mathcal{O}(K^3I^3)$. Typically, we have $K\sim I \ll G $ ($G \gg \mathcal{O}(I^3)$ in most cases), indicating that $G$ (the number of plane waves) is significantly larger than both $K$ (the number of k points) and $I$ (the number of orbitals). Additionally, the FFT operations entail a complexity similar to other plane wave calculations, specifically $\mathcal{O}(KIG\log G)$, due to the need for efficient transformations of energy integrals in reciprocal space. 
    

    
    \newpage
    \section{Numerical Results}
    \label{sec:results}
    In this section we provide evidence for our core claims and evaluate the efficacy of the proposed approach by focusing on four key questions:
    \begin{enumerate}
    \item \textbf{Self-diagonalization}. Can our method diagonalize the Kohn-Sham Hamiltonian as  theoretically expected, even without explicitly applying any diagonalization operation?
    \item \textbf{Fermi-Dirac distribution}. Can the algorithm produce the correct Fermi-Dirac distribution of the occupation numbers?
    \item \textbf{Band Structure}. Is it possible to obtain the same electronic band structure with this approach as with conventional self-consistent Kohn-Sham DFT?
    \item \textbf{Scaling}. How does the scaling compare to other direct optimization methods that handle occupations?
    \end{enumerate}

    To answer these questions regarding the proposed method, we implement it in Python and test it on representative crystal structures, including aluminum and silicon. Our implementation utilizes the automatic differentiation framework JAX \cite{jax2018github}. All experiments are conducted on a single NVIDIA A100 GPU with 40GB of memory. We use the local-density approximation exchange (LDA\_X) as our functional is all cases.\cite{slater1972self}

    \subsection{Self-diagonalization} \label{sect:self-diagonalization}

\begin{figure}
    \centering
    
    \includegraphics[width=1\textwidth]{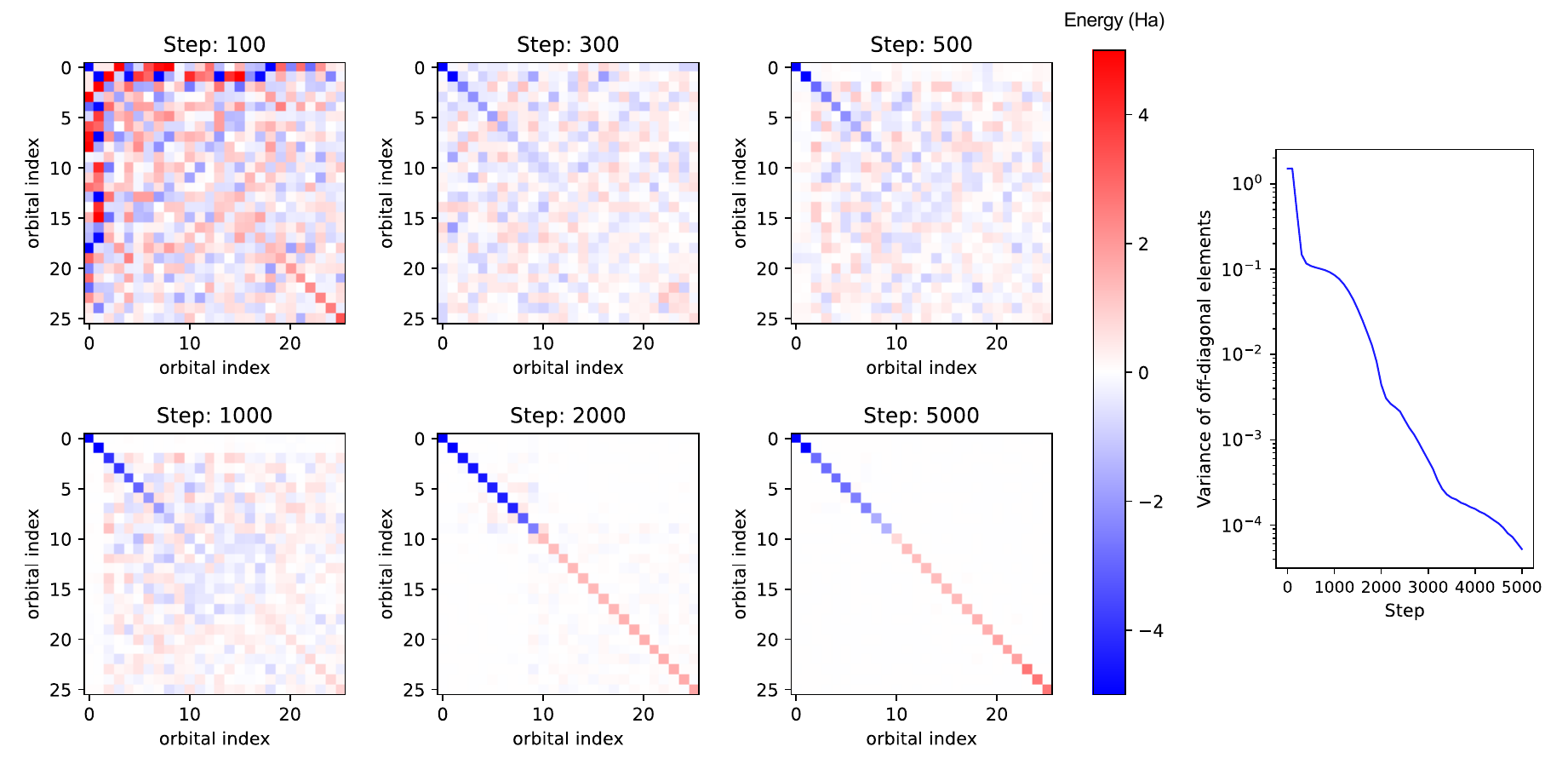}
    \caption{The self-diagonalization process of the Kohn-Sham Hamiltonian matrix $\hat h$ for an aluminum crystal at the $\Gamma$ point, at a temperature of $T=0.01$ Ha.  
    Left: A visualization of the Hamiltonian matrix at steps 100, 300, 500, 1000, 2000 and 5000. The color represents the matrix elements of $\hat h$: blue indicates values below zero down to $-5$ Ha, red indicates values above zero up to $5$ Ha, and white indicates  close to zero, as represented in the colorbar. The plot on the right shows how the variance of the off-diagonal elements of $\hat h$ decreases as the optimization progresses. 
    It is evident that as the free energy (Eq. \eqref{eq:free_energy}) is minimized, $\hat h$ gradually self-diagonalizes.}
    \label{fig:al-self-diag}
    \vspace{0.4in}
\end{figure}

    In this test, we minimize the free energy as defined in Eq. \eqref{eq:free_energy} for the face-centered cubic (FCC) aluminum crystal.\cite{mulder2010hydrogen, vaitkus2023workflow} The self-diagonalization process of the Kohn-Sham Hamiltonian matrix $\hat h$ (defined in Eq. \eqref{eq:hamil_matrix}) for such crystal at a finite temperature (T=0.01 Ha) is shown in Figure \ref{fig:al-self-diag}. The colors in the figure represent the energy in Hartrees: blue indicates values below zero, red indicates values above zero, and white indicates zero. The optimization consists of 5000 steps, and we use the Adam optimizer \cite{kingma2014adam}. As the optimization progresses, $\hat h$ gradually becomes diagonal, clearly seen in the final frame of 5000 steps in which a colored diagonal is surrounded by white (very close to zero) off-diagonal matrix elements. Note that the computation of the full $\hat h$, which we have presented here for illustration, is not required in the running of our algorithm. At any time the Hamiltonian matrix can be constructed from the orbital coefficients via Eqs. \eqref{eq:hamil_matrix} and \eqref{eq:bloch} if desired.


\subsection{Fermi-Dirac Distribution}

 \begin{figure}
    \centering
    \includegraphics[width=0.95\textwidth]{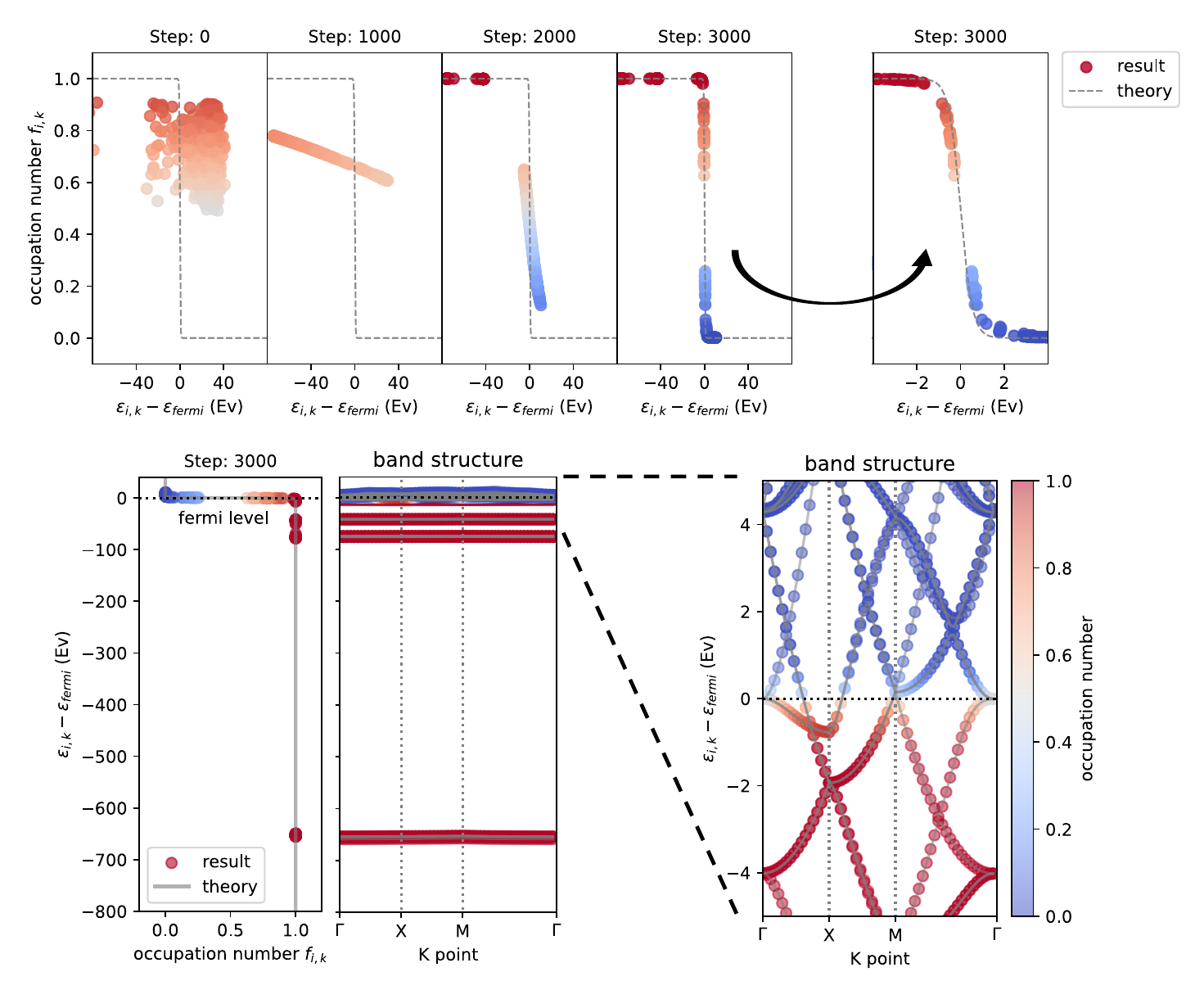}
    \caption{A visualization of the change of occupation numbers during the optimization for a FCC aluminum crystal. Top: Occupation numbers as a function of Hamiltonian diagonal matrix elements relative to the Fermi level ($\varepsilon_i(\kv) - \varepsilon_{\text{fermi}}$). Each point represents a diagonal element-occupation number pair for a potentially occupied orbital. The color indicates the value of the occupation number, with red representing 1 and blue representing 0. A theoretical Fermi-Dirac distribution is shown as a dotted line. The occupation number distributions are displayed at steps 0, 1000, 2000, and 3000. The rightmost figure focuses on the eigenvalues near the Fermi level within a narrow energy range (x-axis) at step 3000. Bottom: An illustration of the band structure of a metal (aluminum based on an FCC conventional unit-cell) and its relation to the occupation numbers.}
     \label{fig:fermi-dirac}
\end{figure}
In this test, we investigate the evolution of the occupation number distribution and its relation to the band structure, using  aluminum as an example.   For the energy calculation, we employ a $64\times 64 \times 64$ FFT mesh, and a $3\times 3 \times 3$ $\kv$-point mesh, at a temperature $T=0.01Ha$, along with a temperature annealing scheme to enhance convergence. The optimization process is depicted in Figure \ref{fig:fermi-dirac}. In the figure, each solid circle represents an occupation number and a diagonal element of the hamiltonian matrix, which will become a Kohn-Sham eigenvalue at the end of the optimization process. The gray dotted curve illustrates the theoretical Fermi-Dirac distribution. The Fermi level is determined by fitting a Fermi-Dirac function to the occupation numbers, with the Fermi level corresponding to the fitted Fermi-Dirac function equals $0.5$. Initially, at step 0, the variational parameters $\Xv_\kv$ and $\Yv$ are randomly initialized, causing the occupation numbers to be scattered around 0.5. As iterations progress, the occupation numbers gradually converge to the theoretical Fermi-Dirac distribution, verifying that our approach can yield accurate Fermi-Dirac distributions of occupation numbers associated with  Kohn-Sham eigenvalues as we progressively optimize the free energy. Additionally, we present the band structure of FCC aluminum, noting that the bands are reordered across the k-points and are partially occupied where they cross the Fermi level, which is essential for determining the thermodynamic properties of the material.

%

%
%
%
%
\subsection{Band Structure}

\begin{figure}[tb]
    \centering
    \begin{subfigure}[b]{0.45\textwidth}
        \centering
        \includegraphics[width=\textwidth]{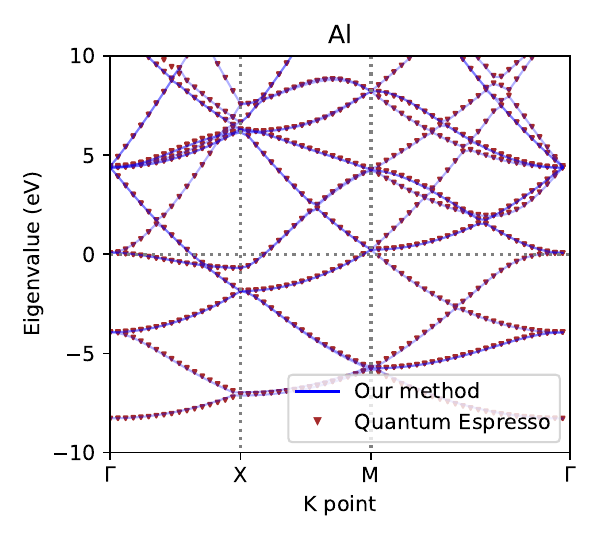}
    \end{subfigure}
    \hspace{-0.1in}
    \begin{subfigure}[b]{0.45\textwidth}
        \centering
        \includegraphics[width=\textwidth]{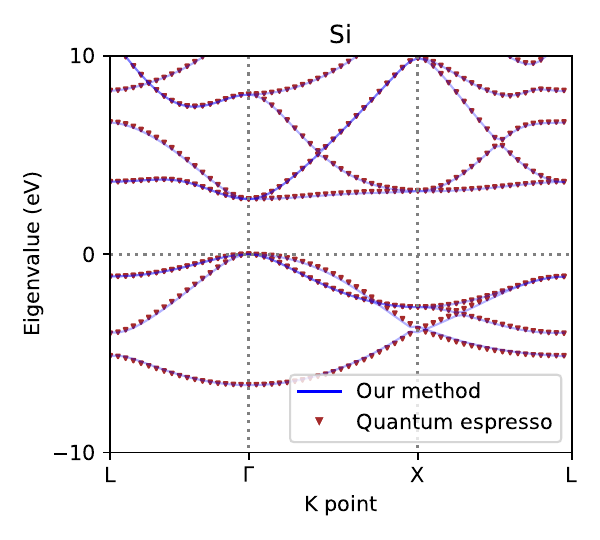}
    \end{subfigure}
\caption{Comparison of the electronic band structures of Aluminum and Silicon calculated using the proposed method and Quantum Espresso. We use $3\times 3 \times 3$ k-point mesh, cutoff energy of $100$ Ha, smearing/temperature values of $T=0.01Ha$ for both materials. As our method is an all-electron method, we have tuned the pseudo-potentials in our Quantum Espresso calculations to allow for an all-electron calculation. Quantum Espresso employs the conventional self-consistent field (SCF) Kohn-Sham DFT with LDA\_X and a plane-wave basis in the pseudopotential projector-augmented wave formalism. The pseudopotentials used in Quantum Espresso are modified such that all-electron calculations are conducted to match our method for evaluation purposes.}
\label{fig:bandstructure}
\vspace{0.4in}
\end{figure}

In this test we compare our computed band structure of Aluminum and Silicon\cite{wyckoff, vaitkus2023workflow} results to those computed the conventional self-consistent (SCF) method, as implemented in Quantum Espresso.\cite{giannozzi2009quantum} The resulting band structures are presented in Figure \ref{fig:bandstructure}. The two methods yield  similar energy band structures, indicating a strong agreement between the two approaches. This demonstrates the accuracy of the proposed method in reproducing the correct band structures and suggests that the occupation numbers and eigenfunctions are also well-aligned with the conventional SCF method. 


\subsection{Scaling comparison}

\begin{figure}[tb]
    \centering
    \includegraphics[width=0.6\textwidth]{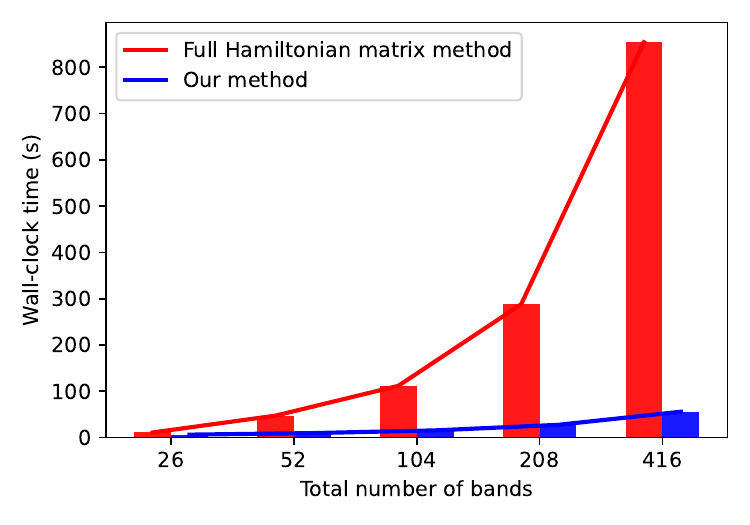}
    \caption{Scaling comparison between the proposed method and full Hamiltonian matrix method. The x-axis represents the number of orbitals (dimension of the Hamiltonian matrix), while the y-axis shows the total training time of 1000 iterations.}
    \label{fig:scaling}
\end{figure}

In this experiment, we compare the scaling performance of our method against the full Hamiltonian matrix method. The full Hamiltonian matrix method encompasses a series of studies that address occupation at finite temperature through direct optimization \cite{marzari1997ensemble, freysoldt2009direct, ivanov2021direct, ruiz2013variational}. In our implementation, we compute the full Hamiltonian matrix $\hat h$ at each step, allowing us to calculate the free energy through Eqs. \eqref{Omega2}, \eqref{Energy}, and \eqref{Entropy}. The primary goal of this experiment is to demonstrate that our approach, which relies solely on the diagonal elements of the Hamiltonian matrix, offers a significant efficiency advantage. We conduct our calculations on an aluminum crystal using a $2 \times 2 \times 2$ $\kv$-point mesh and a $48\times 48 \times 48$ FFT mesh. We vary the total number of bands, including extra empty bands, from 26 to 416. In Figure \ref{fig:scaling} we present our results. The X-axis represents the number of bands, corresponding to the dimension of the Hamiltonian matrix, while the Y-axis shows the total wall-clock execution time (in seconds) for 1000 iterations. As expected, the wall-clock execution time increases with the number of bands for both methods. However, our method consistently exhibits a significant reduction in training time compared to the full Hamiltonian matrix method across all tested band counts (26, 52, 104, 208, and 416). The increased efficiency stems from the fact that our method only requires the calculation of the diagonal elements of the Hamiltonian matrix, while the full Hamiltonian matrix methods necessitates off-diagonal information, resulting in one order of magnitude longer time. This indicates that our method scales more efficiently with  increasing  dimension of the Hamiltonian matrix.

\newpage
\section{Conclusion}

In this work we have presented a new direct optimization approach for the calulation of eigenfunctions and occupations numbers for Kohn-Sham DFT. There are several beneficial features of our approach, specifically our method:
\begin{enumerate}
    \item is an unconstrained search problem with a single iterative step as all physical constraints have been incorporated into the parameterisation of the orbitals and the occupations;
    \item does not require eigenvalue decomposition operations as by requiring that our occupation matrix is diagonal, it automatically diagonalizes the Hamiltonian matrix;
    \item only requires that we compute the diagonal elements of the occupation matrix and the Hamiltonian matrix which significantly reduces the computational time and storage requirements for this part of our calculations;
    \item is fully differentiable, and our implementation in the JAX automatic differentiation framework allows us to take advantage of this in our gradient descent approach.
\end{enumerate}

The most significant impact of this research is that it paves the way for developing other machine learning-based DFT methods. Since this method is fully differentiable and seamlessly integrated with automatic differentiation frameworks widely used in the machine learning community, it greatly simplifies the development of machine learning-assisted density functional theory. This includes tasks such as training neural network-based functionals, potentials, or force fields, allowing for more accurate, efficient, and scalable simulations of complex materials.

\newpage
\section{Acknowledgments}
This research is supported by the Ministry of Education, Singapore, under its Research Centre of Excellence award to the Institute for Functional Intelligent Materials Project No. EDUNC-33-18-279-V12.

\clearpage
\bibliography{ref}

\providecommand{\latin}[1]{#1}
\makeatletter
\providecommand{\doi}
  {\begingroup\let\do\@makeother\dospecials
  \catcode`\{=1 \catcode`\}=2 \doi@aux}
\providecommand{\doi@aux}[1]{\endgroup\texttt{#1}}
\makeatother
\providecommand*\mcitethebibliography{\thebibliography}
\csname @ifundefined\endcsname{endmcitethebibliography}
  {\let\endmcitethebibliography\endthebibliography}{}
\begin{mcitethebibliography}{44}
\providecommand*\natexlab[1]{#1}
\providecommand*\mciteSetBstSublistMode[1]{}
\providecommand*\mciteSetBstMaxWidthForm[2]{}
\providecommand*\mciteBstWouldAddEndPuncttrue
  {\def\EndOfBibitem{\unskip.}}
\providecommand*\mciteBstWouldAddEndPunctfalse
  {\let\EndOfBibitem\relax}
\providecommand*\mciteSetBstMidEndSepPunct[3]{}
\providecommand*\mciteSetBstSublistLabelBeginEnd[3]{}
\providecommand*\EndOfBibitem{}
\mciteSetBstSublistMode{f}
\mciteSetBstMaxWidthForm{subitem}{(\alph{mcitesubitemcount})}
\mciteSetBstSublistLabelBeginEnd
  {\mcitemaxwidthsubitemform\space}
  {\relax}
  {\relax}

\bibitem[Mrovec and Berger(2021)Mrovec, and Berger]{mrovec2021diagonalization}
Mrovec,~M.; Berger,~J. A diagonalization-free optimization algorithm for
  solving Kohn--Sham equations of closed-shell molecules. \emph{Journal of
  Computational Chemistry} \textbf{2021}, \emph{42}, 492--504\relax
\mciteBstWouldAddEndPuncttrue
\mciteSetBstMidEndSepPunct{\mcitedefaultmidpunct}
{\mcitedefaultendpunct}{\mcitedefaultseppunct}\relax
\EndOfBibitem
\bibitem[Ivanov \latin{et~al.}(2021)Ivanov, Levi, Jonsson, and
  J{\'o}nsson]{ivanov2021method}
Ivanov,~A.~V.; Levi,~G.; Jonsson,~E.~O.; J{\'o}nsson,~H. Method for calculating
  excited electronic states using density functionals and direct orbital
  optimization with real space grid or plane-wave basis set. \emph{Journal of
  Chemical Theory and Computation} \textbf{2021}, \emph{17}, 5034--5049\relax
\mciteBstWouldAddEndPuncttrue
\mciteSetBstMidEndSepPunct{\mcitedefaultmidpunct}
{\mcitedefaultendpunct}{\mcitedefaultseppunct}\relax
\EndOfBibitem
\bibitem[Levi \latin{et~al.}(2020)Levi, Ivanov, and
  J{\'o}nsson]{levi2020variational}
Levi,~G.; Ivanov,~A.~V.; J{\'o}nsson,~H. Variational density functional
  calculations of excited states via direct optimization. \emph{Journal of
  Chemical Theory and Computation} \textbf{2020}, \emph{16}, 6968--6982\relax
\mciteBstWouldAddEndPuncttrue
\mciteSetBstMidEndSepPunct{\mcitedefaultmidpunct}
{\mcitedefaultendpunct}{\mcitedefaultseppunct}\relax
\EndOfBibitem
\bibitem[Levi \latin{et~al.}(2020)Levi, Ivanov, and
  J{\'o}nsson]{levi2020variational2}
Levi,~G.; Ivanov,~A.~V.; J{\'o}nsson,~H. Variational calculations of excited
  states via direct optimization of the orbitals in DFT. \emph{Faraday
  Discussions} \textbf{2020}, \emph{224}, 448--466\relax
\mciteBstWouldAddEndPuncttrue
\mciteSetBstMidEndSepPunct{\mcitedefaultmidpunct}
{\mcitedefaultendpunct}{\mcitedefaultseppunct}\relax
\EndOfBibitem
\bibitem[Yao and Su(2024)Yao, and Su]{yao2024enhancing}
Yao,~Y.-F.; Su,~N.~Q. Enhancing reduced density matrix functional theory
  calculations by coupling orbital and occupation optimizations. \emph{The
  Journal of Physical Chemistry A} \textbf{2024}, \relax
\mciteBstWouldAddEndPunctfalse
\mciteSetBstMidEndSepPunct{\mcitedefaultmidpunct}
{}{\mcitedefaultseppunct}\relax
\EndOfBibitem
\bibitem[Schlegel and McDouall(1991)Schlegel, and McDouall]{schlegel1991you}
Schlegel,~H.~B.; McDouall,~J. Do you have SCF stability and convergence
  problems? \emph{Computational advances in organic chemistry: Molecular
  structure and reactivity} \textbf{1991}, 167--185\relax
\mciteBstWouldAddEndPuncttrue
\mciteSetBstMidEndSepPunct{\mcitedefaultmidpunct}
{\mcitedefaultendpunct}{\mcitedefaultseppunct}\relax
\EndOfBibitem
\bibitem[Daniels and Scuseria(2000)Daniels, and
  Scuseria]{daniels2000converging}
Daniels,~A.~D.; Scuseria,~G.~E. Converging difficult SCF cases with conjugate
  gradient density matrix search. \emph{Physical Chemistry Chemical Physics}
  \textbf{2000}, \emph{2}, 2173--2176\relax
\mciteBstWouldAddEndPuncttrue
\mciteSetBstMidEndSepPunct{\mcitedefaultmidpunct}
{\mcitedefaultendpunct}{\mcitedefaultseppunct}\relax
\EndOfBibitem
\bibitem[Jorgensen and Hart(2021)Jorgensen, and
  Hart]{jorgensen2021effectiveness}
Jorgensen,~J.~J.; Hart,~G.~L. Effectiveness of smearing and tetrahedron
  methods: best practices in DFT codes. \emph{Modelling and Simulation in
  Materials Science and Engineering} \textbf{2021}, \emph{29}, 065014\relax
\mciteBstWouldAddEndPuncttrue
\mciteSetBstMidEndSepPunct{\mcitedefaultmidpunct}
{\mcitedefaultendpunct}{\mcitedefaultseppunct}\relax
\EndOfBibitem
\bibitem[Ren and Liu(2022)Ren, and Liu]{ren2022impacts}
Ren,~F.; Liu,~F. Impacts of polarizable continuum models on the SCF convergence
  and DFT delocalization error of large molecules. \emph{The Journal of
  Chemical Physics} \textbf{2022}, \emph{157}\relax
\mciteBstWouldAddEndPuncttrue
\mciteSetBstMidEndSepPunct{\mcitedefaultmidpunct}
{\mcitedefaultendpunct}{\mcitedefaultseppunct}\relax
\EndOfBibitem
\bibitem[Marzari \latin{et~al.}(1997)Marzari, Vanderbilt, and
  Payne]{marzari1997ensemble}
Marzari,~N.; Vanderbilt,~D.; Payne,~M.~C. Ensemble density-functional theory
  for ab initio molecular dynamics of metals and finite-temperature insulators.
  \emph{Physical review letters} \textbf{1997}, \emph{79}, 1337\relax
\mciteBstWouldAddEndPuncttrue
\mciteSetBstMidEndSepPunct{\mcitedefaultmidpunct}
{\mcitedefaultendpunct}{\mcitedefaultseppunct}\relax
\EndOfBibitem
\bibitem[Marzari \latin{et~al.}(1999)Marzari, Vanderbilt, De~Vita, and
  Payne]{marzari1999thermal}
Marzari,~N.; Vanderbilt,~D.; De~Vita,~A.; Payne,~M. Thermal contraction and
  disordering of the Al (110) surface. \emph{Physical review letters}
  \textbf{1999}, \emph{82}, 3296\relax
\mciteBstWouldAddEndPuncttrue
\mciteSetBstMidEndSepPunct{\mcitedefaultmidpunct}
{\mcitedefaultendpunct}{\mcitedefaultseppunct}\relax
\EndOfBibitem
\bibitem[Ivanov \latin{et~al.}(2021)Ivanov, J{\'o}nsson, Vegge, and
  J{\'o}nsson]{ivanov2021direct}
Ivanov,~A.~V.; J{\'o}nsson,~E.~{\"O}.; Vegge,~T.; J{\'o}nsson,~H. Direct energy
  minimization based on exponential transformation in density functional
  calculations of finite and extended systems. \emph{Computer Physics
  Communications} \textbf{2021}, \emph{267}, 108047\relax
\mciteBstWouldAddEndPuncttrue
\mciteSetBstMidEndSepPunct{\mcitedefaultmidpunct}
{\mcitedefaultendpunct}{\mcitedefaultseppunct}\relax
\EndOfBibitem
\bibitem[Freysoldt \latin{et~al.}(2009)Freysoldt, Boeck, and
  Neugebauer]{freysoldt2009direct}
Freysoldt,~C.; Boeck,~S.; Neugebauer,~J. Direct minimization technique for
  metals in density functional theory. \emph{Physical Review B} \textbf{2009},
  \emph{79}, 241103\relax
\mciteBstWouldAddEndPuncttrue
\mciteSetBstMidEndSepPunct{\mcitedefaultmidpunct}
{\mcitedefaultendpunct}{\mcitedefaultseppunct}\relax
\EndOfBibitem
\bibitem[Pham and Khaliullin(2024)Pham, and Khaliullin]{pham2024direct}
Pham,~H.~D.; Khaliullin,~R.~Z. Direct Unconstrained Optimization of Molecular
  Orbital Coefficients in Density Functional Theory. \emph{Journal of Chemical
  Theory and Computation} \textbf{2024}, \relax
\mciteBstWouldAddEndPunctfalse
\mciteSetBstMidEndSepPunct{\mcitedefaultmidpunct}
{}{\mcitedefaultseppunct}\relax
\EndOfBibitem
\bibitem[Wu and Van~Voorhis(2005)Wu, and Van~Voorhis]{wu2005direct}
Wu,~Q.; Van~Voorhis,~T. Direct optimization method to study constrained systems
  within density-functional theory. \emph{Physical Review A—Atomic,
  Molecular, and Optical Physics} \textbf{2005}, \emph{72}, 024502\relax
\mciteBstWouldAddEndPuncttrue
\mciteSetBstMidEndSepPunct{\mcitedefaultmidpunct}
{\mcitedefaultendpunct}{\mcitedefaultseppunct}\relax
\EndOfBibitem
\bibitem[Yang \latin{et~al.}(2006)Yang, Meza, and Wang]{yang2006constrained}
Yang,~C.; Meza,~J.~C.; Wang,~L.-W. A constrained optimization algorithm for
  total energy minimization in electronic structure calculations. \emph{Journal
  of Computational Physics} \textbf{2006}, \emph{217}, 709--721\relax
\mciteBstWouldAddEndPuncttrue
\mciteSetBstMidEndSepPunct{\mcitedefaultmidpunct}
{\mcitedefaultendpunct}{\mcitedefaultseppunct}\relax
\EndOfBibitem
\bibitem[Yang \latin{et~al.}(2007)Yang, Meza, and Wang]{yang2007trust}
Yang,~C.; Meza,~J.~C.; Wang,~L.-W. A trust region direct constrained
  minimization algorithm for the Kohn--Sham equation. \emph{SIAM Journal on
  Scientific Computing} \textbf{2007}, \emph{29}, 1854--1875\relax
\mciteBstWouldAddEndPuncttrue
\mciteSetBstMidEndSepPunct{\mcitedefaultmidpunct}
{\mcitedefaultendpunct}{\mcitedefaultseppunct}\relax
\EndOfBibitem
\bibitem[Canc{\`e}s \latin{et~al.}(2021)Canc{\`e}s, Kemlin, and
  Levitt]{cances2021convergence}
Canc{\`e}s,~E.; Kemlin,~G.; Levitt,~A. Convergence analysis of direct
  minimization and self-consistent iterations. \emph{SIAM Journal on Matrix
  Analysis and Applications} \textbf{2021}, \emph{42}, 243--274\relax
\mciteBstWouldAddEndPuncttrue
\mciteSetBstMidEndSepPunct{\mcitedefaultmidpunct}
{\mcitedefaultendpunct}{\mcitedefaultseppunct}\relax
\EndOfBibitem
\bibitem[Grumbach \latin{et~al.}(1994)Grumbach, Hohl, Martin, and
  Car]{grumbach1994ab}
Grumbach,~M.; Hohl,~D.; Martin,~R.; Car,~R. Ab initio molecular dynamics with a
  finite-temperature density functional. \emph{Journal of Physics: Condensed
  Matter} \textbf{1994}, \emph{6}, 1999\relax
\mciteBstWouldAddEndPuncttrue
\mciteSetBstMidEndSepPunct{\mcitedefaultmidpunct}
{\mcitedefaultendpunct}{\mcitedefaultseppunct}\relax
\EndOfBibitem
\bibitem[Ismail-Beigi and Arias(2000)Ismail-Beigi, and Arias]{ismail2000new}
Ismail-Beigi,~S.; Arias,~T. New algebraic formulation of density functional
  calculation. \emph{Computer Physics Communications} \textbf{2000},
  \emph{128}, 1--45\relax
\mciteBstWouldAddEndPuncttrue
\mciteSetBstMidEndSepPunct{\mcitedefaultmidpunct}
{\mcitedefaultendpunct}{\mcitedefaultseppunct}\relax
\EndOfBibitem
\bibitem[Hohenberg and Kohn(1964)Hohenberg, and
  Kohn]{hohenberg1964inhomogeneous}
Hohenberg,~P.; Kohn,~W. Inhomogeneous electron gas. \emph{Physical review}
  \textbf{1964}, \emph{136}, B864\relax
\mciteBstWouldAddEndPuncttrue
\mciteSetBstMidEndSepPunct{\mcitedefaultmidpunct}
{\mcitedefaultendpunct}{\mcitedefaultseppunct}\relax
\EndOfBibitem
\bibitem[Mermin(1965)]{mermin1965thermal}
Mermin,~N.~D. Thermal properties of the inhomogeneous electron gas.
  \emph{Physical Review} \textbf{1965}, \emph{137}, A1441\relax
\mciteBstWouldAddEndPuncttrue
\mciteSetBstMidEndSepPunct{\mcitedefaultmidpunct}
{\mcitedefaultendpunct}{\mcitedefaultseppunct}\relax
\EndOfBibitem
\bibitem[Gonze \latin{et~al.}(2024)Gonze, Rostami, and
  Tantardini]{gonze2024variational}
Gonze,~X.; Rostami,~S.; Tantardini,~C. Variational density functional
  perturbation theory for metals. \emph{Phys. Rev. B} \textbf{2024},
  \emph{109}, 014317\relax
\mciteBstWouldAddEndPuncttrue
\mciteSetBstMidEndSepPunct{\mcitedefaultmidpunct}
{\mcitedefaultendpunct}{\mcitedefaultseppunct}\relax
\EndOfBibitem
\bibitem[Janak(1978)]{janak1978proof}
Janak,~J.~F. Proof that $\frac{\partial E}{\partial n_i} = \varepsilon$ in
  density-functional theory. \emph{Physical Review B} \textbf{1978}, \emph{18},
  7165\relax
\mciteBstWouldAddEndPuncttrue
\mciteSetBstMidEndSepPunct{\mcitedefaultmidpunct}
{\mcitedefaultendpunct}{\mcitedefaultseppunct}\relax
\EndOfBibitem
\bibitem[Landau and Lifshitz(1964)Landau, and Lifshitz]{landaulifshitz}
Landau,~L.~D.; Lifshitz,~E.~M. \emph{Statistical Physics: Volume 5 of Course of
  Theoretical Physics}; Institute of Physical Problems, U.S.S.R. Academy of
  Sciences: Moscow, 1964\relax
\mciteBstWouldAddEndPuncttrue
\mciteSetBstMidEndSepPunct{\mcitedefaultmidpunct}
{\mcitedefaultendpunct}{\mcitedefaultseppunct}\relax
\EndOfBibitem
\bibitem[Breuer and Petruccione(2002)Breuer, and Petruccione]{breuer2002theory}
Breuer,~H.-P.; Petruccione,~F. \emph{The theory of open quantum systems};
  Oxford University Press, USA, 2002\relax
\mciteBstWouldAddEndPuncttrue
\mciteSetBstMidEndSepPunct{\mcitedefaultmidpunct}
{\mcitedefaultendpunct}{\mcitedefaultseppunct}\relax
\EndOfBibitem
\bibitem[Methfessel and Paxton(1989)Methfessel, and Paxton]{methfessel1989high}
Methfessel,~M.; Paxton,~A. High-precision sampling for Brillouin-zone
  integration in metals. \emph{physical review B} \textbf{1989}, \emph{40},
  3616\relax
\mciteBstWouldAddEndPuncttrue
\mciteSetBstMidEndSepPunct{\mcitedefaultmidpunct}
{\mcitedefaultendpunct}{\mcitedefaultseppunct}\relax
\EndOfBibitem
\bibitem[dos Santos and Marzari(2023)dos Santos, and Marzari]{dos2023fermi}
dos Santos,~F.~J.; Marzari,~N. Fermi energy determination for advanced smearing
  techniques. \emph{Physical Review B} \textbf{2023}, \emph{107}, 195122\relax
\mciteBstWouldAddEndPuncttrue
\mciteSetBstMidEndSepPunct{\mcitedefaultmidpunct}
{\mcitedefaultendpunct}{\mcitedefaultseppunct}\relax
\EndOfBibitem
\bibitem[Fattebert(2022)]{fattebert2022robust}
Fattebert,~J.-L. A robust solver for wavefunction-based density functional
  theory calculations. \emph{Electronic Structure} \textbf{2022}, \emph{4},
  015002\relax
\mciteBstWouldAddEndPuncttrue
\mciteSetBstMidEndSepPunct{\mcitedefaultmidpunct}
{\mcitedefaultendpunct}{\mcitedefaultseppunct}\relax
\EndOfBibitem
\bibitem[Mortensen \latin{et~al.}(2024)Mortensen, Larsen, Kuisma, Ivanov,
  Taghizadeh, Peterson, Haldar, Dohn, Sch{\"a}fer, J{\'o}nsson, \latin{et~al.}
  others]{mortensen2024gpaw}
Mortensen,~J.~J.; Larsen,~A.~H.; Kuisma,~M.; Ivanov,~A.~V.; Taghizadeh,~A.;
  Peterson,~A.; Haldar,~A.; Dohn,~A.~O.; Sch{\"a}fer,~C.;
  J{\'o}nsson,~E.~{\"O}., \latin{et~al.}  GPAW: An open Python package for
  electronic structure calculations. \emph{The Journal of Chemical Physics}
  \textbf{2024}, \emph{160}\relax
\mciteBstWouldAddEndPuncttrue
\mciteSetBstMidEndSepPunct{\mcitedefaultmidpunct}
{\mcitedefaultendpunct}{\mcitedefaultseppunct}\relax
\EndOfBibitem
\bibitem[Roberts and Roberts(2020)Roberts, and Roberts]{roberts2020qr}
Roberts,~D.~A.; Roberts,~L.~R. QR and LQ Decomposition Matrix Backpropagation
  Algorithms for Square, Wide, and Deep--Real or Complex--Matrices and Their
  Software Implementation. \emph{arXiv preprint arXiv:2009.10071}
  \textbf{2020}, \relax
\mciteBstWouldAddEndPunctfalse
\mciteSetBstMidEndSepPunct{\mcitedefaultmidpunct}
{}{\mcitedefaultseppunct}\relax
\EndOfBibitem
\bibitem[Bradbury \latin{et~al.}(2018)Bradbury, Frostig, Hawkins, Johnson,
  Leary, Maclaurin, Necula, Paszke, Vander{P}las, Wanderman-{M}ilne, and
  Zhang]{jax2018github}
Bradbury,~J.; Frostig,~R.; Hawkins,~P.; Johnson,~M.~J.; Leary,~C.;
  Maclaurin,~D.; Necula,~G.; Paszke,~A.; Vander{P}las,~J.;
  Wanderman-{M}ilne,~S.; Zhang,~Q. {JAX}: composable transformations of
  {P}ython+{N}um{P}y programs. 2018; \url{http://github.com/google/jax}\relax
\mciteBstWouldAddEndPuncttrue
\mciteSetBstMidEndSepPunct{\mcitedefaultmidpunct}
{\mcitedefaultendpunct}{\mcitedefaultseppunct}\relax
\EndOfBibitem
\bibitem[Paszke \latin{et~al.}(2019)Paszke, Gross, Massa, Lerer, Bradbury,
  Chanan, Killeen, Lin, Gimelshein, Antiga, \latin{et~al.}
  others]{paszke2019pytorch}
Paszke,~A.; Gross,~S.; Massa,~F.; Lerer,~A.; Bradbury,~J.; Chanan,~G.;
  Killeen,~T.; Lin,~Z.; Gimelshein,~N.; Antiga,~L., \latin{et~al.}  Pytorch: An
  imperative style, high-performance deep learning library. \emph{Advances in
  neural information processing systems} \textbf{2019}, \emph{32}\relax
\mciteBstWouldAddEndPuncttrue
\mciteSetBstMidEndSepPunct{\mcitedefaultmidpunct}
{\mcitedefaultendpunct}{\mcitedefaultseppunct}\relax
\EndOfBibitem
\bibitem[Liu and Nocedal(1989)Liu, and Nocedal]{liu1989limited}
Liu,~D.~C.; Nocedal,~J. On the limited memory BFGS method for large scale
  optimization. \emph{Mathematical programming} \textbf{1989}, \emph{45},
  503--528\relax
\mciteBstWouldAddEndPuncttrue
\mciteSetBstMidEndSepPunct{\mcitedefaultmidpunct}
{\mcitedefaultendpunct}{\mcitedefaultseppunct}\relax
\EndOfBibitem
\bibitem[Duchi \latin{et~al.}(2011)Duchi, Hazan, and Singer]{duchi2011adaptive}
Duchi,~J.; Hazan,~E.; Singer,~Y. Adaptive subgradient methods for online
  learning and stochastic optimization. \emph{Journal of machine learning
  research} \textbf{2011}, \emph{12}\relax
\mciteBstWouldAddEndPuncttrue
\mciteSetBstMidEndSepPunct{\mcitedefaultmidpunct}
{\mcitedefaultendpunct}{\mcitedefaultseppunct}\relax
\EndOfBibitem
\bibitem[Kingma(2014)]{kingma2014adam}
Kingma,~D.~P. Adam: A method for stochastic optimization. \emph{arXiv preprint
  arXiv:1412.6980} \textbf{2014}, \relax
\mciteBstWouldAddEndPunctfalse
\mciteSetBstMidEndSepPunct{\mcitedefaultmidpunct}
{}{\mcitedefaultseppunct}\relax
\EndOfBibitem
\bibitem[Das \latin{et~al.}(2024)Das, Agarwal, Sanghavi, and
  Dhillon]{das2024towards}
Das,~R.; Agarwal,~N.; Sanghavi,~S.; Dhillon,~I.~S. Towards Quantifying the
  Preconditioning Effect of Adam. \emph{arXiv preprint arXiv:2402.07114}
  \textbf{2024}, \relax
\mciteBstWouldAddEndPunctfalse
\mciteSetBstMidEndSepPunct{\mcitedefaultmidpunct}
{}{\mcitedefaultseppunct}\relax
\EndOfBibitem
\bibitem[Slater and Johnson(1972)Slater, and Johnson]{slater1972self}
Slater,~J.~C.; Johnson,~K.~H. Self-consistent-field X $\alpha$ cluster method
  for polyatomic molecules and solids. \emph{Physical Review B} \textbf{1972},
  \emph{5}, 844\relax
\mciteBstWouldAddEndPuncttrue
\mciteSetBstMidEndSepPunct{\mcitedefaultmidpunct}
{\mcitedefaultendpunct}{\mcitedefaultseppunct}\relax
\EndOfBibitem
\bibitem[Mulder \latin{et~al.}(2010)Mulder, Assfour, Huot, Dingemans,
  Wagemaker, and Ramirez-Cuesta]{mulder2010hydrogen}
Mulder,~F.~M.; Assfour,~B.; Huot,~J.; Dingemans,~T.~J.; Wagemaker,~M.;
  Ramirez-Cuesta,~A. Hydrogen in the metal- organic framework Cr MIL-53.
  \emph{The Journal of Physical Chemistry C} \textbf{2010}, \emph{114},
  10648--10655\relax
\mciteBstWouldAddEndPuncttrue
\mciteSetBstMidEndSepPunct{\mcitedefaultmidpunct}
{\mcitedefaultendpunct}{\mcitedefaultseppunct}\relax
\EndOfBibitem
\bibitem[Vaitkus \latin{et~al.}(2023)Vaitkus, Merkys, Sander, Quir{\'o}s,
  Thiessen, Bolton, and Gra{\v{z}}ulis]{vaitkus2023workflow}
Vaitkus,~A.; Merkys,~A.; Sander,~T.; Quir{\'o}s,~M.; Thiessen,~P.~A.;
  Bolton,~E.~E.; Gra{\v{z}}ulis,~S. A workflow for deriving chemical entities
  from crystallographic data and its application to the Crystallography Open
  Database. \emph{Journal of Cheminformatics} \textbf{2023}, \emph{15},
  123\relax
\mciteBstWouldAddEndPuncttrue
\mciteSetBstMidEndSepPunct{\mcitedefaultmidpunct}
{\mcitedefaultendpunct}{\mcitedefaultseppunct}\relax
\EndOfBibitem
\bibitem[Wyckoff(1964)]{wyckoff}
Wyckoff,~R. W.~G. \emph{Crystal Structures}; Wiley Interscience: New York,
  1964\relax
\mciteBstWouldAddEndPuncttrue
\mciteSetBstMidEndSepPunct{\mcitedefaultmidpunct}
{\mcitedefaultendpunct}{\mcitedefaultseppunct}\relax
\EndOfBibitem
\bibitem[Giannozzi \latin{et~al.}(2009)Giannozzi, Baroni, Bonini, Calandra,
  Car, Cavazzoni, Ceresoli, Chiarotti, Cococcioni, Dabo, \latin{et~al.}
  others]{giannozzi2009quantum}
Giannozzi,~P.; Baroni,~S.; Bonini,~N.; Calandra,~M.; Car,~R.; Cavazzoni,~C.;
  Ceresoli,~D.; Chiarotti,~G.~L.; Cococcioni,~M.; Dabo,~I., \latin{et~al.}
  QUANTUM ESPRESSO: a modular and open-source software project for quantum
  simulations of materials. \emph{Journal of physics: Condensed matter}
  \textbf{2009}, \emph{21}, 395502\relax
\mciteBstWouldAddEndPuncttrue
\mciteSetBstMidEndSepPunct{\mcitedefaultmidpunct}
{\mcitedefaultendpunct}{\mcitedefaultseppunct}\relax
\EndOfBibitem
\bibitem[Ruiz-Serrano and Skylaris(2013)Ruiz-Serrano, and
  Skylaris]{ruiz2013variational}
Ruiz-Serrano,~{\'A}.; Skylaris,~C.-K. A variational method for density
  functional theory calculations on metallic systems with thousands of atoms.
  \emph{The Journal of chemical physics} \textbf{2013}, \emph{139}\relax
\mciteBstWouldAddEndPuncttrue
\mciteSetBstMidEndSepPunct{\mcitedefaultmidpunct}
{\mcitedefaultendpunct}{\mcitedefaultseppunct}\relax
\EndOfBibitem
\end{mcitethebibliography}

\end{document}